\documentclass[twocolumn,aps,showpacs,floats,superscriptaddress]{revtex4}
\usepackage[latin9]{inputenc}
\usepackage{amsmath}
\usepackage{amssymb}
\usepackage{graphicx}

\makeatletter

\providecommand{\tabularnewline}{\\}

\@ifundefined{textcolor}{}
{%
 \definecolor{BLACK}{gray}{0}
 \definecolor{WHITE}{gray}{1}
 \definecolor{RED}{rgb}{1,0,0}
 \definecolor{GREEN}{rgb}{0,1,0}
 \definecolor{BLUE}{rgb}{0,0,1}
 \definecolor{CYAN}{cmyk}{1,0,0,0}
 \definecolor{MAGENTA}{cmyk}{0,1,0,0}
 \definecolor{YELLOW}{cmyk}{0,0,1,0}
}

\usepackage{color}

\def\NOT(#1,#2){\OneQubitGate(#1,#2){$X$}}

\makeatother

\begin{document}

\title{Experimental implementation of quantum gates through actuator qubits}

\author{Jingfu Zhang}

\affiliation{Fakultät Physik, Technische Universität Dortmund, D-44221 Dortmund,
Germany }

\author{Daniel Burgarth}

\affiliation{Department of Mathematics and Physics, Aberystwyth University, SY23
3BZ Aberystwyth, United Kingdom}

\author{Raymond Laflamme}

\affiliation{\textit{{Institute for Quantum Computing and Department of Physics,
University of Waterloo, Waterloo, Ontario, Canada N2L 3G1}}}

\affiliation{\textit{Perimeter Institute for Theoretical Physics, Waterloo, Ontario,
N2J 2W9, Canada}}

\author{Dieter Suter}

\affiliation{Fakultät Physik, Technische Universität Dortmund, D-44221 Dortmund,
Germany }

\date{\today}
\begin{abstract}
Universal quantum computation requires the implementation of arbitrary
control operations on the quantum register. In most cases, this is
achieved by external control fields acting selectively on each qubit
to drive single-qubit operations. In combination with a drift Hamiltonian
containing interactions between the qubits, this allows the implementation
of any required gate operation. Here, we demonstrate an alternative
scheme that does not require local control for all qubits: we implement
one- and two-qubit gate operations on a set of target qubits indirectly,
through a combination of gates on directly controlled actuator qubits
with a drift Hamiltonian that couples actuator and target qubits.
Experiments are performed on nuclear spins, using radio-frequency
pulses as gate operations and magnetic-dipole couplings for the drift
Hamiltonian.
\end{abstract}


\maketitle Techniques for controlling quantum systems
\cite{RevModPhys.79.53,RevModPhys.76.1037,QCrew10} have been
developed in various fields, such as quantum computing, where
quantum mechanical two-level systems (qubits) are used to store
information and external control fields process the information by
driving quantum gate operations
~\cite{QCrew00,NielsenChuang,Stolze:2008xy}. A general purpose
quantum computer requires that the control operations can
implement all possible logical operations. This can be achieved,
e.g., by generating arbitrary rotations of all qubits and a static
system Hamiltonian that includes interactions between pairs of
qubits~\cite{NielsenChuang,Stolze:2008xy}.

In some cases, this approach is difficult or impossible to
implement. Examples include systems, where some qubits couple
weakly or not at all to external fields, e.g. when qubits are
stored in noiseless or decoherence-free subspaces
\cite{PhysRevLett.79.3306} or in the case of hybrid quantum
registers consisting of electronic and nuclear spins
\cite{JPhys18.S807,nature07295,NVCreview13}. While qubits in
noiseless subsystems do not interact with control fields by
design, the interaction of nuclear spins with control fields is
some four orders of magnitude weaker than that of electronic
spins.  Control operations generated by direct irradiation of
nuclear spins are therefore slow and it might be desired to avoid
them.  A number of recent papers
\cite{coryPRL12,PhysRevA.81.040303,PhysRevA.82.052333,Khaneja07,CoryPRA,BoughPRL}
proposed schemes for implementing quantum control by directly
manipulating only to a small subset of qubits. In the following,
we distinguish between the directly controlled qubits, the
actuator qubits, and the indirectly controlled qubits, to which we
refer as target qubits. Two similar examples were recently
reported for the case of spin systems consisting of an electron
spin as actuator and nuclear spins as target qubits
\cite{CoryPRA,BoughPRL}. Here, we use heteronuclear spin systems,
where one spin species is the actuator subsystem, while the other
species represents the target subsystem. Compared to previous
work, we extend the size of the total quantum register to five
qubits. The dynamical Lie algebra is calculated explicitly first,
to determine to which degree our system is controllable
\cite{quantumcontrol}.


We use two different systems to demonstrate the indirect control approach.
The smaller one consists of one actuator and two target qubits. All
three qubits are nuclear spins and the interactions between them are
magnetic dipole couplings. 
We denote the qubit in the actuator system as qubit 1, and the two
qubits in the target system as qubits 2 and 3, respectively. The static
Hamiltonian for the whole system is
\begin{equation}
H=H_{A}+H_{T}+H_{AT}\label{HNM}
\end{equation}
where $H_{A}$ refers to the actuator system, $H_{T}$ to the target
system and $H_{AT}$ describes the interaction between them. Their
structure is
\begin{eqnarray}
H_{A} & = & -\pi\nu_{1}Z_{1}\label{HamNMRH}\\
H_{T} & = & -\pi(\nu_{2}Z_{2}+\nu_{3}Z_{3})\nonumber \\
 &  & +\frac{\pi D_{23}}{2}(2Z_{2}Z_{3}-X_{2}X_{3}-Y_{2}Y_{3})\\
H_{AT} & = & \pi(D_{12}Z_{1}Z_{2}+D_{13}Z_{1}Z_{3}).
\end{eqnarray}
Here $X_{i}$, $Y_{i}$, $Z_{i}$ denote Pauli matrices acting on
qubit $i$, $\nu_{i}$ denote the chemical shifts and $D_{ij}$ the
dipolar coupling constants. The control fields are only applied to
qubit 1, so the control Hamiltonian can be written as
\begin{equation}
H_{C}(t)=B_{x}(t)X_{1}+B_{y}(t)Y_{1}.\label{Hc}
\end{equation}
The Lie algebra of the possible control operations on this system
is spanned by the operators that can be generated by repeatedly evaluating
the commutators between the control Hamiltonian $H_{C}$ and the drift
Hamiltonian $H$ \cite{quantumcontrol}. The resulting Lie algebra
includes 22 terms that can be written as
\begin{eqnarray}
 &  & \{X_{1},Y_{1},Z_{1}\}\otimes\{E_{2}E_{3},Z_{2},Z_{3},\label{Liealgebra}\\
 &  & Y_{2}X_{3}-X_{2}Y_{3},X_{2}X_{3}+Y_{2}Y_{3},Z_{2}Z_{3}\},\nonumber \\
 &  & \{E_{1}\}\otimes\{Y_{2}X_{3}-X_{2}Y_{3},X_{2}X_{3}+Y_{2}Y_{3},\nonumber \\
 &  & Z_{2}-Z_{3},-\nu_{2}Z_{2}-\nu_{3}Z_{3}+D_{23}Z_{2}Z_{3}\},\nonumber
\end{eqnarray}
where $E_{k}$ is the unit operator of spin $k$. The Lie algebra
does not include the unit operator $E_{1}E_{2}E_{3}$, since all Hamiltonian
terms (\ref{HNM}-\ref{Hc}) are traceless. Clearly, this allows full
quantum control of the actuator system, but it does not allow full
control of the whole system, which would require $4^{3}=64$ operators.
Nevertheless, it allows the implementation of many useful control
operations in the target system. The interesting terms include the
Dzyaloshinskii-Moriya interaction $Y_{2}X_{3}-X_{2}Y_{3}$, which
is an exchange interaction relevant for some multiferroic materials
\cite{DM1,DM2}, the three-body interaction $Z_{1}Z_{2}Z_{3}$ \cite{corythreebody}
which is a useful resource for implementing time-optimal operations
\cite{PhysRevA.65.032301,PhysRevA.89.042315}, and the $XY$-interaction
$X_{2}X_{3}+Y_{2}Y_{3}$ which allows, e.g., the implementation of
a quantum state transfer along a spin chain \cite{PhysRevLett.92.187902,PhysRevA.72.012331,PhysRevA.76.012317}.

The terms in the set (\ref{Liealgebra}) can be simplified by choosing
specific evolution times. As an example, the last term can become
equivalent to $\nu_{2}Z_{2}+\nu_{3}Z_{3}$ if the evolution time $\tau_{m}$
is chosen such that $\tau_{m}D_{23}=m\pi$, where $m$ is an arbitrary
integer. By combining this with the element $Z_{2}-Z_{3}$, we can
implement qubit-specific $z$-rotations of the target system qubits
\begin{equation}
U_{z,k}(\theta)=e^{i\theta Z_{k}/2}\label{zrotany}
\end{equation}
for certain values of $\theta$. Using optimized control fields $B_{x,y}(t)$,
these gates can be implemented with high fidelity (calculated fidelity
$>$0.99) for both target qubits. The details and the results of the
numerical simulation are given in the supplementary material (SM)
\cite{SM}.

As specific examples, we implement the following operations:
\begin{eqnarray}
U_{z,k}(-\pi/2) & = & e^{-i\frac{\pi}{4}Z_{k}}\label{zrot}\\
U_{z}(-\pi) & = & e^{-i\frac{\pi}{2}(Z_{2}+Z_{3})}\label{zrot2}\\
U_{23}(\theta) & = & e^{i\theta(X_{2}X_{3}+Y_{2}Y_{3})}.\label{zrot3}
\end{eqnarray}
All three gates are important operations for quantum information processing
and the $XY$-interaction can also be used for the transfer of quantum
states. It can be used as a SWAP gate (up to a known phase factor)
by choosing $\theta=\pi/4$, or as an entangling gate, with $\theta=\pi/8$.


\begin{figure}
\centering{}\includegraphics[width=0.7\columnwidth]{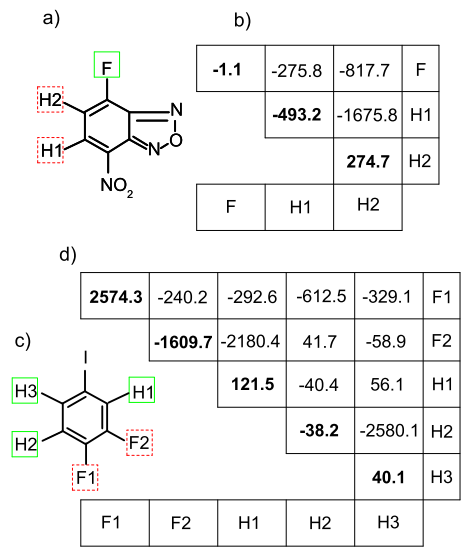} 
\caption{(color online). Structure and Hamiltonian constants of the molecules
used as quantum registers. The actuator and target qubits are marked
by solid and dashed rectangles, respectively. (a) 4-Fluoro-7-nitro-2,1,3-benzoxadiazole.The
fluorine spin F is used as the actuator qubit 1, and the proton spins
H1 and H2 are the target qubits 2 and 3. (b) Hamiltonian parameters
in frequency units (Hz): the diagonal elements are the chemical shifts
in a 11.7 T field, the off-diagonal terms represent the dipolar coupling
constants. (c) 1,2-Difluoro-4-iodobenzene. The proton spins H1 - H3
are the actuator qubits 1 - 3, and the fluorine spins F1 and F2 are
the target qubits 4 and 5. (d) The Hamiltonian parameters of the molecule
(c). \label{MolParam}}
\end{figure}

For the experimental implementation, we choose two molecules dissolved
in nematic liquid crystal solvents as the quantum registers. In the
3-qubit system shown in Fig. \ref{MolParam} (a), we assign the spins
F, H1 and H2 as qubits 1 - 3. The Hamiltonian of this system corresponds
to Eq. (\ref{HNM}) if we neglect scalar couplings, which are significantly
smaller than the dipolar couplings. The measured parameters of the
three qubits are listed in Fig. \ref{MolParam} (b). The control pulses
were generated by the gradient ascent pulse engineering (GRAPE) algorithm
\cite{grape,Colm08}. The calculated fidelities of the operations
are $>0.99$. As a second test system, we used the 5-qubit system
shown in Fig. \ref{MolParam} (c). In this case, the actuator system
consists of the three proton spins denoted as qubits 1 - 3, which
can be fully controlled, and the target system consists of the two
fluorine spins denoted as qubits 4 - 5. Compared with the 3-qubit
system, this molecule contains two additional qubits in the actuator
system while the size of the target system is the same. This larger
system was chosen as a first step on the way to implementing such
control schemes in scalable systems, which require larger numbers
of controlled qubits. As an example, the implementation of quantum
error correction requires at least 5 physical qubits for a perfect
quantum error correction code \cite{RayPRL96,RayPRL01,zhangPRLQEC}.

As in the 3-qubit case, the actuator system of the 5-qubit system
can be fully controlled, i.e. we can implement the set of operations
spanned by $\mathcal{L}_{a}$ = $\{E_{1},X_{1},Y_{1},Z_{1}\}\otimes\{E_{2},X_{2},Y_{2},Z_{2}\}\otimes\{E_{3},X_{3},Y_{3},Z_{3}\}-\{E_{1}E_{2}E_{3}\}$.
Combining this with the drift operator, the full set of operations
that can be applied to the 5-qubit system includes 382 terms. They
can be represented as
\begin{eqnarray}
 &  & \mathcal{L}_{a}\otimes\{E_{4}E_{5},Z_{4},Z_{5},Y_{4}X_{5}-X_{4}Y_{5},\label{Liealgebra5}\\
 &  & X_{4}X_{5}+Y_{4}Y_{5},Z_{4}Z_{5}\},\nonumber \\
 &  & \{E_{1}E_{2}E_{3}\}\otimes\{Y_{4}X_{5}-X_{4}Y_{5},X_{4}X_{5}+Y_{4}Y_{5},\nonumber \\
 &  & Z_{4}-Z_{5},-\nu_{4}Z_{4}-\nu_{5}Z_{5}+D_{45}Z_{4}Z_{5}\},\nonumber
\end{eqnarray}
The result is similar to that of the 3-qubit system: full quantum
control of the actuator system is possible, in combination with similar
operators for the target system. As experimental examples, we implemented
$U_{45}(\theta=\pi/4)$ for both target qubits (the fluorine spins).
The control pulses were generated by the GRAPE algorithm with a theoretical
fidelity $>$0.987, with the contributions of the scalar couplings
included.

\begin{figure}
\centering{}\includegraphics[width=0.95\columnwidth]{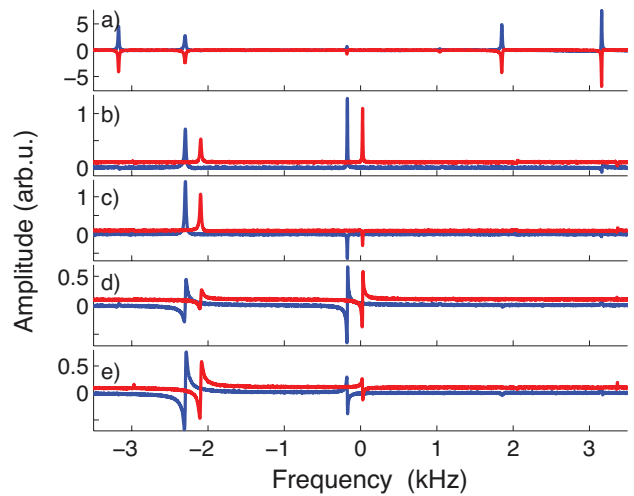}
\caption{(color online). Experimental $^{1}$H-NMR spectra of
4-Fluoro-7-nitro-2,1,3-benzoxadiazole demonstrating the
implementation of quantum gates. The blue curves in (a-e) show
reference spectra for the states $EYE+EEY$, $\mathbf{10}X$,
$\mathbf{1}X\mathbf{0}$, $\mathbf{10}Y$ and
$\mathbf{1}Y\mathbf{0}$, respectively. The red  curves show the
results of the implementation of the gates $U_{z}(-\pi)$ to
$EYE+EEY$, $U_{z,3}(-\pi/2)$ to $-\mathbf{1}\mathbf{0}Y$,
$U_{23}(\pi/4)$ to $-\mathbf{1}\mathbf{0}Y$, $U_{23}(\pi/4)$ to
$\mathbf{1}X\mathbf{0}$ and $U_{z,2}(-\pi/2)$ to
$\mathbf{1}X\mathbf{0}$. In figures (b-e), we shifted the spectra
for the gate implementation for easier comparison with the
reference spectra.\label{figRefsmain}}
\end{figure}

To demonstrate the operation $U_{z}(-\pi)$ of Eq. (\ref{zrot2}),
we applied it to the input state $EYE+EEY$, where we use the abbreviated
notation $ABC=A_{1}\otimes B_{2}\otimes C_{3}$. Following the usual
convention for ensemble quantum computing, we describe the system
by its deviation density matrix, i.e. the traceless part of the density
operator \cite{deviationM}. The input state thus corresponds to the
target qubits oriented along the $y$-axis and the $z$-rotation should
rotate them to the $-y$-axis. The experimental results are shown
in Fig. \ref{figRefsmain} (a). The spectra were obtained by letting
the two states before and after applying the gate operation evolve
under the drift Hamiltonian, measuring the $y$-magnetization of the
system as a function of time and applying a Fourier transformation.
Comparing the two spectra, we find the expected effect that the $U_{z}(-\pi)$
operation inverts the spins and thus the observable resonance lines.
The absolute value of the spectral lines after the inversion is reduced
by $\approx15$\%, to $c=-0.86\pm0.09$. This reduction can be attributed
to relaxation: the transverse relaxation times for H1 and H2 range
from $14$ to $36$ ms, as determined from the width of the resonance
lines. 

For the implementation of the gates $U_{z,k}(-\pi/2)$, 
 and $U_{23}(\theta)$, we chose elements from the set
\begin{equation}
\{\mathbf{10}X,-\mathbf{10}Y,\mathbf{1}X\mathbf{0},-\mathbf{1}Y\mathbf{0}\}\label{observs}
\end{equation}
as input states, were $\mathbf{0}\equiv|0\rangle\langle0|$ and $\mathbf{1}\equiv|1\rangle\langle1|$.
For implementing $U_{45}(\pi/4)$ in the 5-qubit system, we chose
the input states as
\begin{equation}
\{\mathbf{0000}X,-\mathbf{0000}Y,\mathbf{000}X\mathbf{0},-\mathbf{000}Y\mathbf{0}\}.\label{observs-1}
\end{equation}
We prepared these input states using the established techniques developed
in quantum information \cite{NMRpps}.

\begin{figure}[h]
\centering{}\includegraphics[width=0.6\columnwidth]{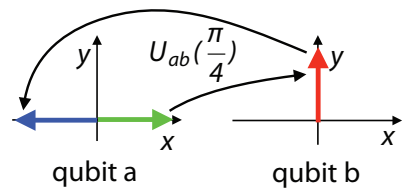} 
\caption{(color online). Graphical representation of the effect of $U_{ab}(\pi/4)$,
which transforms $X_{a}\mathbf{0}_{b}$ $\to$ $\mathbf{0}_{a}Y_{b}$
$\to$ $-X_{a}\mathbf{0}_{b}$ and vice versa, as indicated by the
arrows. The extra qubit in state $\mathbf{0}$ is not shown in the
figure. \label{graphU}}
\end{figure}


The states in set (\ref{observs}) or (\ref{observs-1}) constitute
a basis for the gates, i.e., the set is closed under the effect of
the gates. The states were chosen to give readily observable signals
in NMR spectra as shown in Figs. \ref{figRefsmain} (b - e). 


The single-qubit gates should generate the transformations
$U_{z,a}(-\pi/2):X_{a}\rightarrow Y_{a}\rightarrow-X_{a}$, where
$U:\rho_{A}\rightarrow\rho_{B}$ denotes
$U\rho_{A}U^{\dag}=\rho_{B}$. The two-qubit gates implement
\begin{eqnarray}
U_{ab}(\theta) & : & X_{a}\mathbf{0}_{b}\rightarrow X_{a}\mathbf{0}_{b}\cos(2\theta)+\mathbf{0}_{a}Y_{b}\sin(2\theta)\label{XYrot0X}\\
 &  & Y_{a}\mathbf{0}_{b}\rightarrow Y_{a}\mathbf{0}_{b}\cos(2\theta)-\mathbf{0}_{a}X_{b}\sin(2\theta).\nonumber
\end{eqnarray}
Fig. \ref{graphU} shows a graphical representation of these
transformations. $U_{ab}(\theta=\pi/4)$ is equivalent to the SWAP
gate, up to a phase gate.


The experiment results for the three qubit system are illustrated
by the spectra of Figs. \ref{figRefsmain} (b - e).  To quantify
the performance of the gates, we independently prepared the
predicted final states $\mathbf{10}X$, $\mathbf{1}X\mathbf{0}$,
$\mathbf{10}Y$ and $\mathbf{1}Y\mathbf{0}$ and measured their
spectra. These reference spectra are shown as the blue curves in
Figs. \ref{figRefsmain} (b - e). The red curves represent the
results of the operations. Comparing the amplitudes of the two
spectra in each figure yields the overlap of the state after the
gate implementation with the predicted final state. Table
\ref{tableObs} lists the measured overlaps.

\begin{table}
\centering{}%
\begin{tabular}{|c|c|c|c|c|}
\hline
 & $R_{z,a}(-\frac{\pi}{2})$  & $R_{z,b}(-\frac{\pi}{2})$  & $U_{ab}(\frac{\pi}{4})$  & $U_{ab}(\frac{\pi}{4})$ \tabularnewline
 &  &  & (3-qubits)  & (5-qubits) \tabularnewline
\hline
$\mathbf{0}_{a}X_{b}$  & -  & $0.69\pm0.04$  & $0.67\pm0.02$  & $0.59\pm0.04$ \tabularnewline
\hline
$-\mathbf{0}_{a}Y_{b}$  & -  & $0.68\pm0.04$  & $0.68\pm0.02$  & $0.60\pm0.03$ \tabularnewline
\hline
$X_{a}\mathbf{0}_{b}$  & $0.70\pm0.02$  & -  & $0.58\pm0.04$  & $0.64\pm0.09$ \tabularnewline
\hline
$-Y_{a}\mathbf{0}_{b}$  & $0.71\pm0.02$  & -  & $0.58\pm0.04$  & $0.68\pm0.04$ \tabularnewline
\hline
\end{tabular}\caption{Measured overlaps obtained by fitting the spectra. The input states
are listed in the first column and the gate operations in the first
row. The state of the actuator qubit(s) is omitted. The values in
columns 1-3 were obtained with the 3-qubit system, and the last column
with the 5-qubit system. \label{tableObs}}
\end{table}


\begin{figure}[h]
\centering{}\includegraphics[width=1\columnwidth]{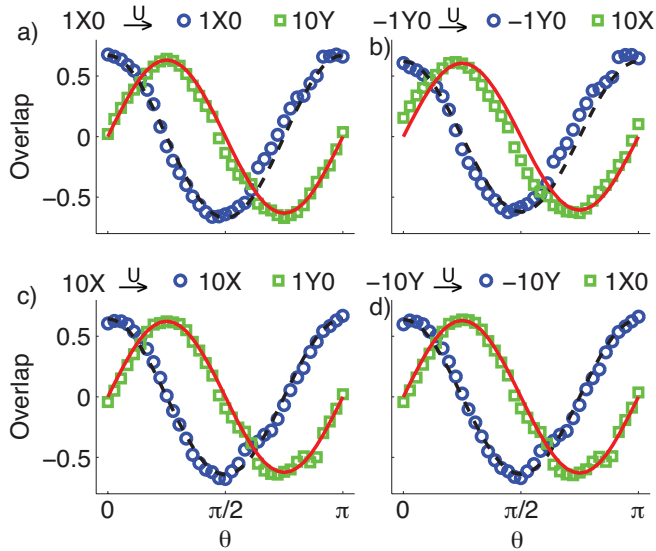} 
\caption{(color online). Experimental results for implementing $U_{23}(\theta)$
in the 3-qubit system for the input states $\mathbf{1}X\mathbf{0}$,
$-\mathbf{1}Y\mathbf{0}$, $\mathbf{10}X$, $-\mathbf{10}Y$, shown
as figures (a-d), respectively. The basis operators for which the
overlaps are determined are given in each panel. The solid and dashed
curves are fits to the experimental data points. 
\label{figXYmain}}
\end{figure}


For an arbitrary angle $\theta$, $U_{23}(\theta)$ transfers the
input state to a linear combination of two states from set (\ref{observs}),
as shown in Eq. (\ref{XYrot0X}). For the experimental data, we determined
the corresponding coefficients by fitting the measured spectra to
a linear combination of the corresponding reference spectra, which
are shown as the blue curves in Figs. \ref{figRefsmain} (b-e). Fig.
\ref{figXYmain} shows the resulting overlap coefficients when $U_{23}(\theta)$
was applied to the four input states in set (\ref{observs}). As a
function of the rotation angle $\theta$, the individual data points
can be fitted to $A_{i}\cos(2\theta)$ and $B_{i}\sin(2\theta)$,
with $A_{1}=0.672\pm0.014$, $A_{2}=0.622\pm0.026$, $A_{3}=0.644\pm0.008$,
$A_{4}=0.639\pm0.007$, and $B_{1}=0.633\pm0.008$, $B_{2}=0.607\pm0.035$,
$B_{3}=0.624\pm0.007$, $B_{4}=0.630\pm0.007$ for the four input
states $\mathbf{1}X\mathbf{0}$, $-\mathbf{1}Y\mathbf{0}$, $\mathbf{10}X$,
$-\mathbf{10}Y$, respectively.

\begin{figure}
\centering{}\includegraphics[width=3.5in]{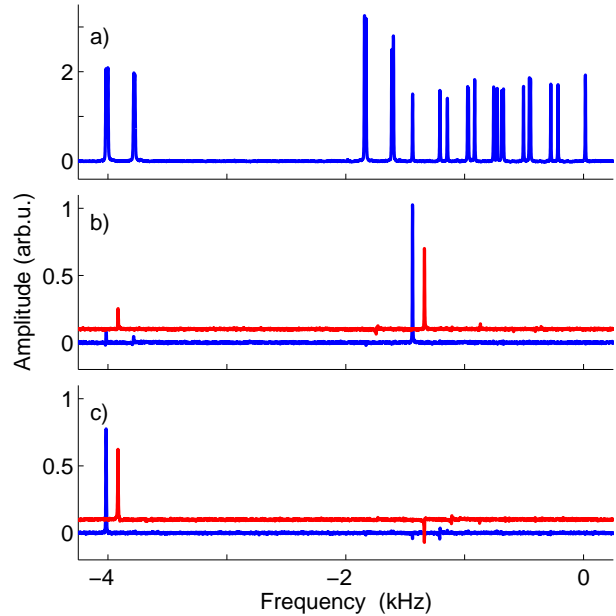} 
\caption{(color online). Experimental $^{19}$F-NMR spectra of the
molecule 1,2-difluoro-4-iodobenzene, demonstrating the performance
of operation $U_{45}(\pi/4)$ in the 5-qubit system. The spectra
represented as the blue curves in figure (a-c) are obtained from
the states $EEEXE+EEEEX$, $\mathbf{000}X\mathbf{0}$, and
$\mathbf{0000}X$, respectively. The spectra represented as the red
curves in figure (b-c) result from the implementation of
$U_{45}(\pi/4)$ to $-\mathbf{0000}Y$ and
$-\mathbf{000}Y\mathbf{0}$.
\label{figswap5}}
\end{figure}


Using a similar strategy, we also implemented $U_{45}(\pi/4)$ in
the 5-qubit system.  Figs. \ref{figswap5} (a-c) show $^{19}$F-NMR
spectra from the states $EEEXE+EEEEX$, $\mathbf{000}X\mathbf{0}$
and $\mathbf{0000}X$ as  blue curves. The red curves in Figs.
\ref{figswap5} (b, c) show the spectra after applying
$U_{45}(\pi/4)$ to $-\mathbf{0000}Y$ and
$-\mathbf{000}Y\mathbf{0}$, respectively. The measured overlaps
from various input states are listed in the last column in Table
\ref{tableObs}.

The main contributions to the imperfections of the gate
implementation can be attributed to (i) finite precision of the
calculated control operations (ii) relaxation and (iii)
experimental errors in the implementation of the gate. We used
numerical simulations of the experiment to quantify these
contributions, as described in the SM (Figs. S3 and S4 and Table
SI). According to these simulations, the fidelity loss for the
gate $U_{45}(\pi/4)$ resulting from (i)-(iii) is $4\%$, $26\%$ and
$8\%$, respectively.


The purpose of this paper was the demonstration that a suitable combination
of local control operations to a subsystem of the total quantum system,
together with a suitable drift Hamiltonian, allows control not only
over the directly controlled qubits (the actuator qubits), but also
partial or full control of the target qubits. For this demonstration,
we used two types of nuclear spins, with one type representing the
actuator qubits, the other the target qubits. The couplings between
the qubits were magnetic dipole interactions. The results show good
agreement between theory and experiment. While these results were
obtained with nuclear spins, the same concept should be applicable
to other systems, such as nitrogen-vacancy centers in diamond \cite{interactionsNVC,interactionsNVC13C},
where the hyperfine interactions provide sufficient resources.

J.Z. acknowledges helpful discussions with J. Filgueiras, and
experimental assistance from M. Holbach and J. Lambert. This work
is supported by the DFG through Su 192/19-1. R.L. thanks CIFAR and
Industry Canada for support.


\end{document}